\documentclass[10pt,aps,pra,notitlepage,nofootinbib,longbibliography,showpacs]{revtex4-1}

\usepackage{amsmath,bm}
\usepackage{graphicx}
\usepackage{hyperref}

\begin{document}

\title{Repulsive long-range forces between anisotropic atoms and dielectrics}
\author{K. V. Shajesh}
\email{shajesh@andromeda.rutgers.edu}
\homepage{http://andromeda.rutgers.edu/~shajesh}
\author{M. Schaden}
\email{mschaden@andromeda.rutgers.edu}
\homepage{http://andromeda.rutgers.edu/~physics/mschaden.htm}
\affiliation{Department of Physics, 
Rutgers, The State University of New Jersey, 
101 Warren Street, Newark, New Jersey - 07102, USA.}

\date{\today}

\begin{abstract}
We investigate long-range forces between atoms with anisotropic
electric polarizability interacting with dielectrics having
anisotropic permittivity in the weak-coupling approximation.
Unstable configurations in which the force between the objects
is repulsive are constructed. Such configurations exist for three
anisotropic atoms as well as for an anisotropic atom above a
dielectric plate with a hole whose permittivity is anisotropic.
Apart from the absolute magnitude of the force, the dependence
on the configuration is qualitatively the same as for metallic
objects for which the anisotropy is a purely geometric effect.
In the weak limit closed analytic expressions for rather
complicated configurations are obtained. The non-monotonic
dependence of the interaction energy on separation is related
to the fact that the electromagnetic Green's dyadic is not
positive definite. The analysis in the weak limit is found to
also semi-quantitatively explain the dependence of Casimir forces
on the orientation of anisotropic dielectrics observed experimentally.
Contrary to the scalar case, irreducible electromagnetic three-body
energies can change sign. We trace this to the fact that the
electromagnetic Green's dyadic is not positive definite.
\end{abstract}

\maketitle

\section{Introduction}

Fifty-seven years after the postulation of attractive long-range
van der Waals forces between neutral atoms \cite{Waals:1873sl},
London \cite{London:1930a,London:1930b,Hettema:2001cq}
described them as a purely quantum mechanical effect. Later
Axilrod and Teller \cite{Axilrod:1943at} and Muto \cite{Muto:1943fc}
independently found that the three-body contribution of three atoms
to the van der Waals-London interaction energy
can change sign and contribute repulsively depending on their configuration.
Repulsive long-range forces have so far been established in
at least two situations\footnote{%
We here do not include the repulsive pressure on a sphere found by
Boyer \cite{Boyer:1968uf} because it does not involve two objects.%
}:
for two dielectric objects with a suitable dielectric 
between them \cite{Dzyaloshinskii:1961fw,Munday:2009fl}, and
for a dielectric object and a magnetic material 
\cite{Feinberg1968pw,Boyer:1969pw}. The latter repulsive effect
has not yet been observed.
Although it was known that the interaction between anisotropic atoms
could have repulsive components \cite{Craig1969ma,Craig:1969pa},
it came as a surprise that the force on a vertically
oriented metallic needle above a hole in a metallic plate
can be repulsive as well \cite{Levin:2010vo}.
The explanation of this effect relied on symmetry arguments
and numerical calculations.
An analytic solution, in the unretarded regime, for a polarizable
particle above a plate with a hole was obtained in \cite{Eberlein:2011hw}
with similar repulsive effects.
By summing two-body contributions in the multiple scattering formalism
the effect could also be verified in the retarded regime \cite{Maghrebi:2011at}.
The origin of such repulsive effects nevertheless does not seem to
have been sufficiently well understood to make reliable predictions
for more general situations. For instance, although very interesting
repulsive configurations were found in \cite{Milton:2011cp,Milton:2011ed}
one of the conclusion was that
``there is no repulsion possible in the weak coupling regime''.
For anisotropic atoms and dielectrics this is not the case.

The interaction between anisotropic atoms has been studied
extensively (see \cite{Babb2005ac} and references therein).
We in the following identify configurations with repulsive
long-range forces between atoms and dielectric materials with
anisotropic polarizabilities and  permittivities. 
For weak polarizability and susceptibility we obtain exact analytic
expressions for a wide range of geometries. The force on an atom
sufficiently close to a dielectric plate with a hole is found to be
repulsive for certain relative orientations of the polarizability
and permeability
of the atom and the dielectric material. It is always attractive
at large separations. The cause for Casimir repulsion is traced to
anisotropies in the polarizability of the objects and to the fact that
the electromagnetic Green's dyadic is not positive definite. For ideal
metals the anisotropy is geometrical and exact results for dilute
dielectrics should qualitatively extend to metals.

We begin by examining long-range forces between anisotropic atoms
and dielectrics. As for scalar Casimir energies 
\cite{Schaden:2011in,Shajesh:2011ec,Shajesh:2011se,Schaden:2011sn}
we can show that the two-body contribution to the electromagnetic
Casimir interaction is always negative, independent of the relative
orientation of polarizability and permittivity tensors. However, contrary
to the scalar case, the electromagnetic Casimir interaction energy in
general is not a monotonic function of separation and some
force-components can change sign. Remarkably, torque-free points at
which the interaction energy does not depend on the orientation of
the atom appear to exist for some configurations of dielectrics.
We then examine the three-body correction to the interaction of
isotropic atoms \cite{Axilrod:1943at,Aub:1960a} and find that it
never dominates two-body contributions to the Casimir energy.
Contrary to the scalar case, irreducible three-body contributions
to the electromagnetic Casimir energy can change sign, because
the electromagnetic Green's dyadic is not positive definite.


\section{Anisotropic polarizable atoms}

Long-range interactions of polarizable atoms are described by the 
van der Waals--London forces \cite{London:1930a,London:1930b}
in the unretarded regime and by Casimir-Polder forces
\cite{Casmir:1947hx} in the retarded regime.
In the multiple scattering formalism the interaction energy for
two polarizable atoms is
\begin{equation}
E_{12} = \frac{1}{2} \int_{-\infty}^{\infty}
\frac{d\zeta}{2\pi}\,\text{Tr}\ln \Big[ {\bf 1} - 
{\bf \Gamma}_0\cdot {\bf T}_1 \cdot {\bf \Gamma}_0\cdot {\bf T}_2 \Big],
\label{E12=Logof}
\end{equation}
an expression already obtained for atoms in \cite{Ninham:1971dp}.
The free Green's dyadic ${\bf \Gamma}_0$ in Eq.~(\ref{E12=Logof}) is
\begin{equation}
{\bf \Gamma}_0({\bf r};i\zeta) = \frac{e^{-|\zeta|r}}{4\pi\,r^3}
\Big[ -u(|\zeta|r) \,{\bf 1} + v(|\zeta| r)\, \hat{\bf r} \hat{\bf r} \Big],
\label{freeGdyadic}
\end{equation}
with $u(x) = 1+x+x^2$, and $v(x) = 3 + 3x + x^2$.
Neglecting quadruple and higher moments,
the scattering matrix ${\bf T}_i$ for the $i$-th atom
with atomic dipole polarizability ${\bm\alpha}_i(i\zeta)$ is 
\begin{equation}
{\bf T}_i({\bf x},{\bf x}^\prime;i\zeta) 
= 4\pi{\bm\alpha}_i(i\zeta)\, \delta^{(3)}({\bf x}-{\bf x}_i)
 \delta^{(3)}({\bf x}^\prime-{\bf x}_i),
\label{Ti-atoms}
\end{equation}
where ${\bf x}_i$ specifies the position of the atom.
The $\delta$-functions in Eq.~(\ref{Ti-atoms}) permit a trivial evaluation
of the spatial integrals of the trace in Eq.~(\ref{E12=Logof}).
For separations $r_{ij}$ satisfying
$r_{ij}^6\gg |{\bm\alpha}_i(0){\bm\alpha}_j(0)|$
the logarithm in Eq.~(\ref{E12=Logof}) may be 
expanded\footnote{Approximating atoms as point-like objects
in Eq.~(\ref{Ti-atoms}) is not justified for
$r_{ij}^6\lesssim |{\bm\alpha}_i(0){\bm\alpha}_j(0)|$.%
}. The weak approximation consists of retaining
only the leading term of this expansion and we have
\begin{eqnarray}
E_{12}^W &=& -\frac{1}{2\pi\, r^6}\int_0^{\infty} d\zeta\, e^{-\zeta r} \Big[ 
u^2(\zeta r)\, \text{Tr}\{{\bm\alpha}_1(i\zeta)\cdot{\bm\alpha}_2(i\zeta)\}
-2u(\zeta r)v(\zeta r)\, \{\hat{\bf r}\cdot{\bm\alpha}_1(i\zeta)\cdot
  {\bm\alpha}_2(i\zeta)\cdot\hat{\bf r}\}
\nonumber \\ && \hspace{32mm}
+ v^2(\zeta r)\, \{\hat{\bf r}\cdot{\bm\alpha}_1(i\zeta)\cdot\hat{\bf r}\}
\{\hat{\bf r}\cdot{\bm\alpha}_2(i\zeta)\cdot\hat{\bf r}\} \Big],
\label{E12W}
\end{eqnarray}
which is the interaction obtained in \cite{Craig1969ma,Craig:1969pa}.
We used the property that polarizability is a symmetric tensor,
of the form ${\bm\alpha}_i=\sum_n \alpha_i^n \hat{\bf e}_i^n \hat{\bf e}_i^n$,
where the $\hat{\bf e}_i^n$, $n=1,2,3$,
are the orthogonal principal axes satisfying
$\hat{\bf e}_i^m\cdot\hat{\bf e}_i^n=\delta^{mn}$ and
$\hat{\bf e}_i^m\times\hat{\bf e}_i^n=\varepsilon^{mnl}\hat{\bf e}_i^l$.
For stable atoms the corresponding principal polarizabilities
$\alpha_i^n$ necessarily are non-negative. A simple model for 
the frequency dependence of the atomic polarizability is
\begin{equation}
{\bm \alpha}_i(i\zeta)={\bm \alpha}_i(0) \frac{\omega_i^2}{\omega_i^2+\zeta^2},
\label{alpha-model}
\end{equation}
where $\omega_i$ is the excitation energy of the (two-level) atom
and ${\bm \alpha}_i(0)$ is its static polarizability.
One of the ways to calculate atomic polarizability is by 
Dalgarno's method \cite{Dalgarno:1955tp}.

The exponential dependence on the separation distance in
Eq.~(\ref{freeGdyadic}) implies that the frequency dependence of
the polarizability ${\bm \alpha_i}(i\zeta)$ in Eq.~(\ref{alpha-model})
is negligible for $\alpha^\frac{1}{3}< c/\omega_i \ll r$.
In this asymptotic retarded (Casimir-Polder) regime the polarizabilities
can be approximated  by their static values and the $\zeta$-integration
in Eq.~(\ref{E12W}) performed to yield \cite{Craig1969ma,Craig:1969pa},
\begin{eqnarray}
E_{12}^\text{CP}({\bm\alpha}_1,{\bm\alpha}_2;{\bf r})
&=& -\frac{1}{8\pi\, r^7} \Big[ 
13\, \text{Tr}\{{\bm\alpha}_1(0)\cdot{\bm\alpha}_2(0)\}
-56\, \{\hat{\bf r}\cdot{\bm\alpha}_1(0)\cdot{\bm\alpha}_2(0)\cdot\hat{\bf r}\}
+ 63\, \{\hat{\bf r}\cdot{\bm\alpha}_1(0)\cdot\hat{\bf r}\}
\{\hat{\bf r}\cdot{\bm\alpha}_2(0)\cdot\hat{\bf r}\}\Big].
\label{E12WCP}
\end{eqnarray}
For atoms with isotropic polarizabilities,
${\bm\alpha}_1=\alpha_1{\bf 1}$ and ${\bm\alpha}_2=\alpha_2{\bf 1}$,
Eq.~(\ref{E12WCP}) reproduces the Casimir-Polder
interaction \cite{Casmir:1947hx}
\begin{equation}
E_{12}^\text{CP}(\alpha_1{\bf 1},\alpha_2{\bf 1};r)
= -\frac{\alpha_1(0)\alpha_2(0)}{r^7} \frac{23}{4\pi}.
\label{E12WCP33}
\end{equation}

In the unretarded (London) regime ($\alpha^\frac{1}{3}< r \ll c/\omega_i$),
the frequency dependence in Eq.~(\ref{freeGdyadic}) may be neglected
and the free dyadic in Eq.~(\ref{E12W}) approximated by the
static dipole-dipole interaction ${\bm\Gamma}_0({\bm r};0)$
to yield \cite{Craig1969ma,Craig:1969pa},
\begin{eqnarray}
E_{12}^\text{Lon} ({\bm\alpha}_1,{\bm\alpha}_2;{\bf r})
&=& -\frac{1}{2\pi\, r^6}\int_0^{\infty}
d\zeta\, \Big[ \text{Tr}\{{\bm\alpha}_1(i\zeta)\cdot{\bm\alpha}_2(i\zeta)\}
-6\, \{ \hat{\bf r}\cdot{\bm\alpha}_1(i\zeta)
  \cdot{\bm\alpha}_2(i\zeta)\cdot\hat{\bf r}\}
\nonumber \\ && \hspace{25mm}
+ 9\, \{ \hat{\bf r}\cdot{\bm\alpha}_1(i\zeta)\cdot\hat{\bf r} \}
\{ \hat{\bf r}\cdot{\bm\alpha}_2(i\zeta)\cdot\hat{\bf r} \} \Big].%
\label{E12WLondon}%
\end{eqnarray}%
For atoms with isotropic polarizabilities this reproduces London's
expression for the van der Waals interaction
\begin{equation}
E_{12}^\text{Lon} (\alpha_1{\bf 1},\alpha_2{\bf 1};r)
= -\frac{3}{\pi\, r^6}\int_0^{\infty}
d\zeta\, \alpha_1(i\zeta)\alpha_2(i\zeta),
\label{E12WLondoniso}
\end{equation}
which is inversely proportional to the sixth power in
the separation $r$.
To evaluate the coefficient, the frequency-dependence of the
polarizabilities has to be known or modeled.
For the simple model in Eq.~(\ref{alpha-model}) 
(letting $\omega_1=\omega_2=\omega_0$) one has 
\begin{equation}
E_{12}^\text{Lon} (\alpha_1{\bf 1},\alpha_2{\bf 1};r)
= -\frac{\alpha_1(0)\alpha_2(0)}{r^6}
\frac{3\omega_0}{4}.
\label{E12WLondoniso-model}
\end{equation}
With the increased computational power it nowadays is only slightly
more complicated to perform the complete frequency integral
in Eq.~(\ref{E12W}) numerically.

To illustrate the orientation dependence of the Casimir-Polder interaction,
consider the special case where one of the atoms is isotropic and the
other a linear molecule that essentially can only be polarized along
its axis with a polarizability given by
${\bm\alpha}_1= \alpha_1\, \hat{\bf e}\, \hat{\bf e}$.
With $\hat{\bf e}\cdot \hat{\bf r}=\cos\theta$,
Eq.~(\ref{E12WCP}) then becomes 
\begin{equation}
E_{12}^\text{CP}({\bm\alpha}_1,\alpha_2{\bf 1};{\bf r})
= -\frac{\alpha_1(0)\alpha_2(0)}{8\pi\, r^7} \Big[ 13 +7 \cos^2\theta \Big].
\label{E12WCP13}
\end{equation}
The minimum energy configuration is at $\theta=0$ or $\theta=\pi$,
corresponding to the alignment of the axis $\hat{\bf e}$
of the molecule with ${\bm r}$.
Note that the interaction energy of Eq.~(\ref{E12WCP13})
is negative for any orientation.

\begin{figure}
\includegraphics{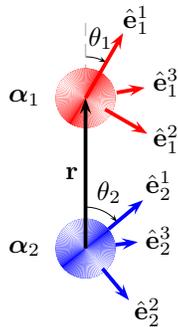}
\caption{Two atoms with anisotropic polarizabilities.}
\label{unisouniso-fig}
\end{figure}

Consider next two atoms with general static anisotropic polarizability tensors
${\bm\alpha}_i=\sum_n \alpha_i^n \hat{\bf e}_i^n \hat{\bf e}_i^n$.
(see Fig.~\ref{unisouniso-fig}.) Inserting these in Eq.~(\ref{E12WCP}),
the interaction energy in the retarded regime is of the form
\begin{equation}
E_{12}^\text{CP}({\bm\alpha}_1,{\bm\alpha}_2;{\bf r}) = -\frac{1}{8\pi\, r^7}
\sum_{m,n=1}^{3} \alpha_1^m(0)\alpha_2^n(0) C_{12}^{mn},
\label{E12WCP11}
\end{equation}
with
\begin{equation}
C_{12}^{mn} = \frac{1}{13} \big| 13 (\hat{\bf e}_1^m\cdot \hat{\bf e}_2^n)
- (28 + i\sqrt{35}) (\hat{\bf e}_1^m\cdot \hat{\bf r}) 
(\hat{\bf e}_2^n\cdot \hat{\bf r}) \big|^2.
\end{equation}
\begin{center} \begin{table}
\begin{tabular}{ccccc}
\hline 
\hspace{3mm} $(\hat{\bf e}^m_1\cdot\hat{\bf e}^n_1)$ 
\hspace{3mm} & \hspace{3mm}
$(\hat{\bf e}_1^m\cdot \hat{\bf r}) (\hat{\bf e}_2^n\cdot\hat{\bf r})$
\hspace{3mm} & \hspace{3mm} $-C_{12}^{mn}$ \hspace{3mm} & \hspace{3mm}
Comments \hspace{3mm} & \hspace{3mm} Example \hspace{3mm} \\
\hline
$\pm 1$ & $\pm 1$ & -20 & Minima & \begin{minipage}{25mm}
\includegraphics[height=10mm]{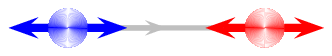} 
\end{minipage} \\
$\pm 1$ & 0 & -13 & Saddle & \begin{minipage}{25mm}
\includegraphics[height=10mm]{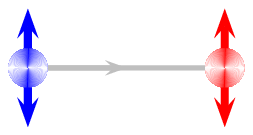} 
\end{minipage} \\
$\pm 1$ & $\pm \frac{4}{9}$ & $-\frac{5}{9}$ & Saddle & \begin{minipage}{25mm}
\includegraphics[height=10mm]{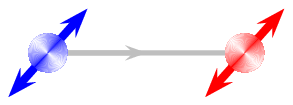}
\end{minipage} \\
0 & 0 & 0 & Maxima & \begin{minipage}{25mm}
\includegraphics[height=10mm]{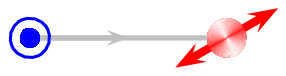}
\end{minipage} \\
\hline
\end{tabular}
\caption{Configurations with extremal energies for two
linearly polarizable molecules. In the rows corresponding to
$C_{12}^{mn}=20$, and $5/9$,
the signs in the first two columns should concur.}
\label{extrema-table}
\end{table} \end{center}
The potential energy surface of the interaction given in Eq.~(\ref{E12WCP11})
is rich, with (local) extrema of individual $C_{12}^{mn}$
at $C_{12}^{mn}=20,13,5/9$, and $0$.
These extremums are summarized in Table~\ref{extrema-table}.
The $C_{12}^{mn}$'s are non-negative and vanish only if
$\hat{\bf e}^m_1\cdot\hat{\bf e}^n_1=0$ and
$(\hat{\bf e}_1^m\cdot \hat{\bf r}) (\hat{\bf e}_2^n\cdot \hat{\bf r})=0$.
Since the eigenvalues of the polarization tensor are non-negative,
the two-body interaction between two anisotropic atoms therefore
is always negative. The maximum value of $C_{12}^{mn}=20$
corresponds to the energetically most favorable orientation,
$\hat{\bf e}_1^m=\pm \hat{\bf r}$, $\hat{\bf e}_2^n=\pm \hat{\bf r}$,
where both atoms are aligned with ${\bf r}$.
The potential energy surface has a saddle point when 
$(\hat{\bf e}_1^m\cdot \hat{\bf r}) (\hat{\bf e}_2^n\cdot \hat{\bf r})=0$
and $\hat{\bf e}^m_1\cdot\hat{\bf e}^n_1=\pm 1$ that corresponds 
to $C_{12}^{mn}=13$. Interesting extremum for
$(\hat{\bf e}_1^m\cdot \hat{\bf r}) = \pm 2/3 =
(\hat{\bf e}_2^n\cdot \hat{\bf r})$
and $\hat{\bf e}^m_1\cdot\hat{\bf e}^n_1=\pm 1$
corresponds to $C_{12}^{mn}=5/9$.
Note that $(\hat{\bf e}_1^m\cdot \hat{\bf r}) = \pm 2/3$ represents
an angle of about $48.2^\circ$ between the polarizabilities 
and $\hat{\bf r}$. The interaction energy in Eq.~(\ref{E12WCP11})
gives rise to a non-central force between anisotropic atoms,
\begin{equation}
{\bf F} = -\frac{1}{2\pi\, r^8}
\sum_{m,n=1}^{3} \alpha_1^m(0)\alpha_2^n(0)
\Big[ 7\, \hat{\bf r}\, C_{12}^{mn} - r {\bm\nabla} C_{12}^{mn} \Big].
\label{force-torque}
\end{equation}
The second term in Eq.~(\ref{force-torque}) is a torque that
vanishes for the extremal configurations of Table ~\ref{extrema-table}.


\section{Casimir repulsion}

\begin{figure}
\includegraphics{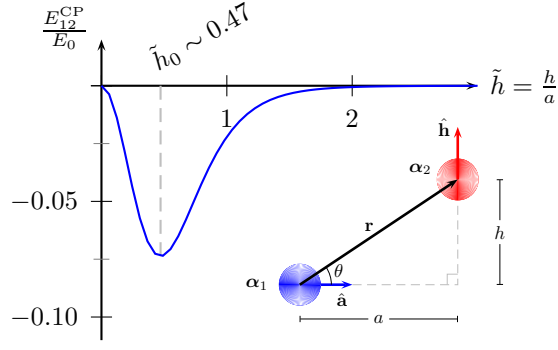}
\caption{Dimensionless energy,
with $E_0=63\alpha_1(0)\alpha_2(0)/8\pi\,a^7$,
for two atoms with orthogonal polarizabilities in Eq.~(\ref{E12WCPortho}),
sketched in the insert, plotted with respect to $h/a$.}
\label{two-ortho-plot-fig}
\end{figure}

Since the Casimir-Polder energy can vanish for particular orientations
of anisotropic atoms and is always negative, it is clear from the
foregoing discussion that it in general is not a monotonic function of the
distance between the atoms and that for fixed orientation of the atoms
components of the force between them can be repulsive.
To see this more explicitly consider two linear atoms with
polarizabilities that are orthogonal to each other:
${\bm\alpha}_1=\alpha_1 \hat{\bf a} \hat{\bf a}$,
${\bm\alpha}_2=\alpha_2 \hat{\bf h} \hat{\bf h}$, with 
$\hat{\bf a} \cdot\hat{\bf h}=0$.
Let $\hat{\bf a} \cdot\hat{\bf r}=\cos\theta$,
$\hat{\bf h} \cdot\hat{\bf r}=\sin\theta$, $r^2=a^2+h^2$.
(See insert in Fig.~\ref{two-ortho-plot-fig}.)
Eq.~(\ref{E12WCP}) gives the interaction energy for this configuration as 
\begin{equation}
E_{12}^\text{CP}({\bm\alpha}_1,{\bm\alpha}_2;{\bf r})
= -\frac{\alpha_1(0)\alpha_2(0)}{a^7} \frac{63}{8\pi}
\frac{\tilde h^2}{(1+\tilde h^2 )^\frac{11}{2}}.
\label{E12WCPortho}
\end{equation}
For fixed horizontal separation $a$, the dimensionless interaction energy
of Eq.~(\ref{E12WCPortho}) is shown in Fig.~\ref{two-ortho-plot-fig}.
It vanishes for $h=0$ and $h\rightarrow \infty$ and is negative otherwise.
The minimum at $\tilde h=h/a=\sqrt{2}/3\sim0.47$ implies that the vertical
component of the force on atom 2 along $\bf{\bf h}$ changes sign
when $3h=\sqrt{2}a$. Regimes with repulsive components of the non-central
force between anisotropic atoms exist for all
values of $\hat{\bf a}\cdot \hat{\bf h}$.
For $|\hat{\bf a}\cdot \hat{\bf h}|$ sufficiently close to unity,
the repulsive regime splits into two or more disjoint regions in $\tilde h$.
In particular, for $|\hat{\bf a}\cdot \hat{\bf h}|=1$ the two
repulsive regimes are disjoint, symmetrical, and above and below the plate.

\begin{figure}
\includegraphics{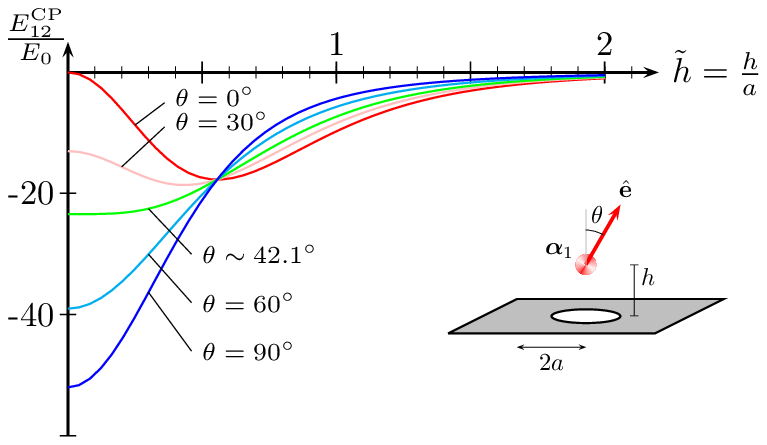}
\hspace{10mm}
\includegraphics{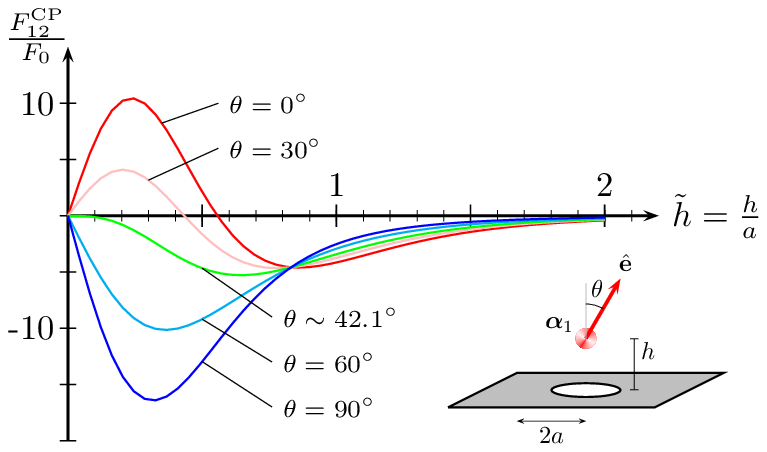}
\caption{The dimensionless energy in Eq.~(\ref{energy-atom-plate-hole}) with
$E_0=\alpha\lambda/320\pi a^5$ (left)
and the dimensionless force in Eq.~(\ref{force-atom-plate-hole}) with
$F_0=\alpha\lambda/64\pi a^5$ (right) as functions of $h/a$
for an anisotropic atom at a height $h$ above the center of a
dielectric plate with a hole of radius $a$ as sketched in the inserts.
The curves correspond to different orientation $\theta$
of the atomic polarizability.} 
\label{atom-on-hole-dielectric-force-fig}
\end{figure}

In the previous example, the horizontal component of the force between
the two atoms was always attractive. However, this component of the force
on atom~2 may be compensated by placing a third atom with polarization tensor
${\bm\alpha}_3= \pm\alpha_1 \hat{\bf a} \hat{\bf a}$ in the symmetric point
with the three atoms forming an isosceles triangle.
Of course, real atoms would not maintain these orientations and
positions unless they are part of a material. These considerations
bring us to the analog of the geometry of a metallic needle above a
metallic plate \cite{Levin:2010vo}. We therefore consider an atom with
anisotropic polarizability centered above a dilute dielectric 
material with a hole as sketched in the inserts of
Fig.~\ref{atom-on-hole-dielectric-force-fig}.
In the weak limit the scattering matrix of dilute dielectric objects
can be approximated by their dielectric permittivity with
${\bf T}_i({\bf x},{\bf y})
\sim {\bm V}_i({\bf x}) \delta^{(3)}({\bf x}-{\bf y})$
and 
${\bm V}_i({\bf x}) =[{\bm \varepsilon}({\bf x})-{\bf 1}] 
\sim 4\pi {\bm\alpha}_i n_i({\bf x})$,
where $n_i({\bf x})$ is the number density of the atoms the dielectric
is composed of and ${\bm\alpha}_i$ is their polarizability. 
In this weak limit Eq.~(\ref{E12=Logof}) in the retarded regime takes the form
\begin{eqnarray}
E_{12}^\text{CP} &=& - \frac{1}{128\,\pi^3} \int d^3x \int d^3x^\prime
\frac{1}{|{\bf x}-{\bf x}^\prime|^7}
\Big[ 13\, \text{Tr} \left\{ {\bf V}_1({\bf x}) 
 \cdot {\bf V}_2({\bf x}^\prime) \right\}
-56\, \hat{\bf r} \cdot {\bf V}_1({\bf x}) \cdot
 {\bf V}_2({\bf x}^\prime) \cdot \hat{\bf r}
\nonumber \\ && \hspace{50mm}
+ 63\, \left\{ \hat{\bf r} \cdot {\bf V}_1({\bf x}) \cdot \hat{\bf r} \right\}
 \left\{ \hat{\bf r} \cdot {\bf V}_2({\bf x}^\prime) \cdot \hat{\bf r} \right\}
\Big].
\label{E12W-V1V2}
\end{eqnarray}
Eq.~(\ref{E12W-V1V2}) generalizes the corresponding expression for 
isotropic dielectric functions in \cite{Milton:2008lw} to the anisotropic case. 

Using Eq.~(\ref{E12W-V1V2}) we now can study the analog of the
configuration proposed in 
\cite{Levin:2010vo} for dilute anisotropic dielectrics and atoms.
In a Cartesian coordinate system with axes in the
$\hat{\bf x}$, $\hat{\bf y}$, and $\hat{\bf z}$ directions, we consider
a dielectric plate in the $xy$-plane with a hole of radius $a$ centered
at the origin and an anisotropic (linear) atom at the height $h$
on the $z$-axis. The scattering potential of the anisotropic atom
is modeled by, 
${\bf V}_1({\bf x})=4\pi {\bm\alpha} \delta(x)\delta(y)\delta(z-h)$
with the anisotropic static polarizability
${\bm\alpha}=\alpha\,\hat{\bf e}\, \hat{\bf e}$ and 
$\hat{\bf e} \cdot \hat{\bf z} = \cos\theta $.
The scattering potential of the dielectric plate of thickness $d$
with a hole of radius $a$ is described by
\begin{eqnarray}
{\bf V}_2({\bf x}) 
&=& ({\bm \varepsilon}-{\bf 1}) \,\theta (\rho -a) [\theta (z+d) - \theta(z)]
\sim {\bm \lambda}\, \theta (\rho -a) \delta (z),
\end{eqnarray}
with  $\rho^2=x^2+y^2$. To simplify the calculation we consider
a thin plate and define 
${\bm \lambda} \sim ({\bm \varepsilon}-{\bf 1}) d
\sim 4\pi {\bm \alpha} \sigma $,
with $\sigma=nd$ the planar density of atoms.
We furthermore assume that the dielectric material is polarizable
in the plane of the dielectric only and therefore demand that
${\bm \lambda} \cdot \hat{\bf z} =0$.
If ${\bm \lambda}$ is isotropic in the $xy$ plane we can insert
${\bm \lambda} =\lambda\, (\hat{\bf x} \hat{\bf x} + \hat{\bf y} \hat{\bf y})$
in Eq.~(\ref{E12W-V1V2}) to obtain
\begin{equation}
E_{12}^\text{CP}(a,h,\theta)
= - \frac{\alpha \lambda}{320\pi\,a^5}
\frac{1}{(1+\tilde h^2)^\frac{9}{2}}
\left[ (36 \tilde h^4 + 97 \tilde h^2 + 26)
+ (4\tilde h^4 + 83 \tilde h^2 - 26) \cos 2\theta \right],
\label{energy-atom-plate-hole}
\end{equation}
for the interaction energy in the retarded regime. 
Fig.~\ref{atom-on-hole-dielectric-force-fig} 
shows the dependence of Eq.~(\ref{energy-atom-plate-hole}) on the
dimensionless height $\tilde h=h/a$ for different orientations
$\theta$ of the anisotropic atom. At the intersection of all the curves in 
Fig.~\ref{atom-on-hole-dielectric-force-fig} the interaction energy
does not depend on the orientation $\theta$ of the atom and the
anisotropic atom is torsion free. This is the case when the coefficient
of $\cos 2\theta$ in Eq.~(\ref{energy-atom-plate-hole}) vanishes,
that is for $4\tilde h^4 + 83 \tilde h^2 - 26=0$,
giving $\tilde h \sim 0.56$. The torsion-free point coincides with
the highest (unstable) equilibrium point, attained for $\theta=0$.
For heights above the torsion-free point the energy is minimized
for vertical orientation of the polarizability and below it the energy is 
minimized for horizontal orientations. The transition from vertical to
horizontal orientation is sudden and happens at the torsion-free
point without expense in energy.
The net force on the atom in $z$-direction is
\begin{equation}
F_{12}^\text{CP}(a,h,\theta)
= - \frac{\alpha \lambda}{64\pi\,a^5}
\frac{\tilde h}{(1+\tilde h^2)^\frac{11}{2}}
\left[ (36 \tilde h^4 + 107 \tilde h^2 + 8)
+ (4\tilde h^4 + 113 \tilde h^2 - 80) \cos 2\theta \right],
\label{force-atom-plate-hole}
\end{equation}
and is shown in Fig.~\ref{atom-on-hole-dielectric-force-fig} as a
function of $\tilde h$ for several orientations $\theta$ of the atom.
The orientation-independent value for the vertical force occurs when
$4\tilde h^4 + 113 \tilde h^2 - 80=0$ and corresponds to $\tilde h \sim 0.83$.
The force on the atom given by Eq.~(\ref{force-atom-plate-hole})
is repulsive when
\begin{equation}
{\tilde h}^2 = \frac{h^2}{a^2} < \frac{-(107+113\cos 2\theta)
+\sqrt{(107+113\cos 2\theta)^2-128(9+\cos 2\theta)(1-10\cos 2\theta)}}
{8(9+\cos 2 \theta)}
\end{equation}
which has real solutions for 
\begin{equation}
0<\theta < \frac{1}{2} \cos^{-1} \left( \frac{1}{10} \right)
\qquad \text{and} \qquad
\pi - \frac{1}{2} \cos^{-1} \left( \frac{1}{10} \right)<\theta < \pi.
\end{equation}
The critical value for the orientation angle,
$\theta_c = 0.5 \cos^{-1}( 1/10) \sim 42.1^\circ$,
is the angle beyond which no repulsive regime exists.

To emphasize how weakly interacting configurations could lead to a 
semi-quantitative understanding of their strongly interacting counterparts,
we next consider the configuration in \cite{Romanowski20081},
where the Casimir force between
gold and the anisotropic cuprate superconductor 
Bi$_2$Sr$_2$CaCu$_2$O$_{8+\delta}$ (BSCCO) was found to decrease in
magnitude by as much as 10-20\% when the BSCCO planes were oriented
parallel to the gold plate rather than perpendicular to it.
Let us replace the gold plate by a semi-infinite dielectric slab with
isotropic permittivity described by
${\bf V}_1({\bf x}) = [{\bm \varepsilon}_1-{\bf 1}] \theta(-z)$
with $({\bm \varepsilon}_1-{\bf 1}) = (\varepsilon_1-1) {\bf 1}$,
and replace the BSCCO by a semi-infinite dielectric slab with 
${\bf V}_2({\bf x}) = [{\bm \varepsilon}_2-{\bf 1}] \theta(z-a)$
and
$({\bm \varepsilon}_2-{\bf 1}) =(\varepsilon_2-1)_\perp\, 
(\hat{\bf x} \hat{\bf x} + \hat{\bf y} \hat{\bf y})
+ (\varepsilon_2-1)_{||} \hat{\bf z} \hat{\bf z}$
when the BSCCO-planes are parallel to the $xy$-plane
(perpendicular cleave in \cite{Romanowski20081}) and 
$({\bm \varepsilon}_2-{\bf 1}) =(\varepsilon_2-1)_{||} \hat{\bf x} \hat{\bf x}
+ (\varepsilon_2-1)_\perp\, (\hat{\bf y} \hat{\bf y} +\hat{\bf z}\hat{\bf z})$
if they are oriented parallel to the $xz$-plane
(parallel cleave in \cite{Romanowski20081}).
Using Eq.~(\ref{E12W-V1V2}) one obtains for the ratio of
the Casimir force in the two orientations
\begin{equation}
\frac{F_\text{$||$ cleave}}{F_\text{$\perp$ cleave}} =1 +  
\frac{2\left[ (\varepsilon_2-1)_\perp - (\varepsilon_2-1)_{||} \right]}
{\left[ 14 (\varepsilon_2-1)_\perp + 9 (\varepsilon_2-1)_{||} \right]}.
\end{equation}
If we assume $(\varepsilon_2-1)_{||} \ll (\varepsilon_2-1)_\perp$,
which is a good approximation for BSCCO, the
magnitude of the Casimir force in the retarded regime changes by
$1/7\sim 14\%$, in semi-quantitative agreement with \cite{Romanowski20081}.


\section{Three-atom Casimir energies} 

Irreducible many-body Casimir energies were proven to remain finite in 
\cite{Schaden:2011in,Schaden:2011sn} when some, but not all, objects overlap.
For a massless scalar quantum field with local potential interactions,
the sign of the $N$-body Casimir energy was found to be simply
$(-1)^{N+1}$. These theorems were verified in all examples studied
in \cite{Schaden:2011in,Shajesh:2011ec,Shajesh:2011se,Schaden:2011sn}
where closed expressions for irreducible
many-body Casimir energies were derived in the framework of the
multiple scattering expansion. The  proof of the sign of irreducible
many-body contributions in the scalar case relied on the positivity
of the free scalar Green's function.
The electromagnetic free Green's dyadic of Eq.~(\ref{freeGdyadic}) 
is not a positive definite operator. The eigenvalues of this matrix,
corresponding to eigenvectors parallel to ${\bf r}$ and orthogonal to it,
are proportional to $(u(x)-v(x),u(x),u(x))$, 
where $u(x)$ and $v(x)$ were defined after Eq.~(\ref{freeGdyadic}).
Here we investigate consequences of this for the sign of the
irreducible three-body contribution to the Casimir energy
between three isotropic atoms \cite{Axilrod:1943at,Aub:1960a}.

\begin{figure}
\includegraphics{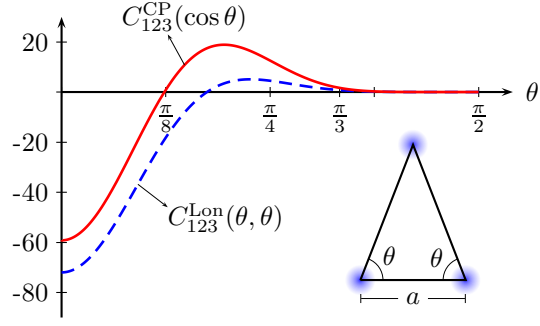}
\caption{Angular dependence of three-body contributions to the
long-range forces in the retarded (CP) and unretarded (Lon)
regime for three atoms forming an isosceles triangle.}
\label{three-body-isosceles-fig}
\end{figure}

Inserting the free Green's dyadic of Eq.~(\ref{freeGdyadic})
into the expression for the irreducible three-body contribution to
the Casimir energy in \cite{Shajesh:2011ec}, and using scattering matrices
for isotropic atoms given by Eq.~(\ref{Ti-atoms}) with 
${\bm \alpha}_i = \alpha_i(i\zeta) {\bf 1}$ one obtains
\begin{eqnarray}
E_{123}^W &=& \frac{1}{\pi} \frac{1}{r_{12}^3r_{23}^3r_{31}^3}
\int_0^\infty d\zeta\,e^{-\zeta (r_{12}+r_{23}+r_{31})}
\alpha_1(i\zeta) \alpha_2(i\zeta) \alpha_3(i\zeta)
\, Q_{123}({\bf r}_{12},{\bf r}_{23}),
\label{3iso-atoms}
\end{eqnarray}
where the displacements ${\bf r}_{ij}$ are the sides of the triangle
formed by the three atoms and satisfy 
${\bf r}_{12}+{\bf r}_{23}+{\bf r}_{31}=0$.
The function $Q_{123}$ in Eq.~(\ref{3iso-atoms}) is
\begin{eqnarray}
Q_{123}({\bf r}_{12},{\bf r}_{23})
&=& 3 u_{12}u_{23}u_{31}
- u_{12}u_{23}v_{31} - u_{12}v_{23}u_{31} 
- v_{12}u_{23}u_{31} 
+ u_{12}v_{23}v_{31} \cos^2\theta_3
\nonumber \\ &&
+ v_{12}u_{23}v_{31} \cos^2\theta_1 + v_{12}v_{23}u_{31} \cos^2\theta_2
+ v_{12}v_{23}v_{31} \cos\theta_1 \cos\theta_2 \cos\theta_3,
\end{eqnarray}
with $u_{ij}=u(|\zeta|r_{ij})$, $v_{ij}=v(|\zeta|r_{ij})$, the
$\theta_i$'s ($i=1,2,3$) being the internal angles of the triangle.
In the unretarded limit Eq.~(\ref{3iso-atoms}) simplifies to the
expression obtained in \cite{Axilrod:1943at},
\begin{eqnarray}
E_{123}^\text{Lon}
&=& \frac{(1+3\cos\theta_1\cos\theta_2\cos\theta_3)}{r_{12}^3r_{23}^3r_{31}^3}
\frac{3}{\pi} \int_0^\infty d\zeta\, 
\alpha_1(i\zeta) \alpha_2(i\zeta) \alpha_3(i\zeta).
\label{London3}
\end{eqnarray}
Using the simple model in Eq.~(\ref{alpha-model})
for three identical isotropic atoms with
$\omega_1=\omega_2=\omega_3=\omega_0$,
the integral in Eq.~(\ref{London3}) is readily performed to 
yield \cite{Axilrod:1949gr},
\begin{eqnarray}
E_{123}^\text{Lon}
&=& \frac{\alpha_1(0) \alpha_2(0) \alpha_3(0)}{a^{10}} 
\,\omega_0a \,C_{123}^\text{Lon}(\theta_1,\theta_2),
\end{eqnarray}
where the angular dependence is given by \cite{Aub:1960a}
\begin{equation}
C_{123}^\text{Lon}(\theta_1,\theta_2)
=\frac{9}{16}
\frac{\sin^6\theta_3(1+3\cos\theta_1\cos\theta_2\cos\theta_3)}
{\sin^3\theta_1 \sin^3\theta_2},
\end{equation}
with the three sides and angles of the triangle related by the
law of sines. It was noted in \cite{Axilrod:1943at} that
the three-body contribution in the unretarded regime
is negative when the atoms form 
an acute triangle and positive when it is (very) obtuse. For isosceles
triangles this behavior is seen in Fig.~\ref{three-body-isosceles-fig}.
Refs. \cite{Axilrod:1949gr,Axilrod:1951ti,Axilrod:1951gr}
attempted to explain the crystal structures of rare gases by three-body
contributions to the total energy. Although this turned out not to be
possible, three-body effects typically contribute between 10-20\%
to the total energy.

The $\zeta$-integral in Eq.~(\ref{3iso-atoms}) can be performed analytically
in the retarded limit but the result is not very illuminating
and too lengthy to be presented here. In the special case
when the atoms are arranged to form an isosceles triangle
with side lengths $r_{12}=a$,
$r_{23}=r_{31}=r$ and $\theta_1=\theta_2=\theta$,
(see insert in Fig.~\ref{three-body-isosceles-fig})
the retarded interaction is \cite{Aub:1960a}
\begin{equation}
E_{123}^\text{CP}(a,\theta) = \frac{\alpha_1(0) \alpha_2(0)\alpha_3(0)}{a^{10}}
\,C_{123}^\text{CP}(\cos\theta),
\end{equation}
with
\begin{eqnarray}
C_{123}^\text{CP}(x) &=& \frac{1}{4\pi}
\bigg[ \frac{2x}{1+x}\bigg]^7 \Big[ 
7 + 49 x + 611 x^2 +1533 x^3 +868 x^4 -1372 x^5
-1672 x^6 -672 x^7 -96 x^8 \Big].
\end{eqnarray}
For an equilateral triangle it yields 
$E_{123}^\text{CP}(a,\pi/3) 
=\alpha_1(0)\alpha_2(0)\alpha_3(0) 1264/243\pi a^{10}$
in agreement with \cite{Aub:1960a}.
As in the unretarded limit, the angular dependence in the retarded limit
also leads to configurations in which the three-body force contributes
repulsively. Fig.~\ref{three-body-isosceles-fig} in fact shows that
the irreducible three-body contribution to the potential is less
attractive in the retarded regime. However, for isotropic atoms,
the three-body contribution never dominates over two-body
contributions in the regime $\alpha <r^3$ where a point-like description
of the atoms is justified. Evidently, the analysis of atomic
many-body interaction is only a little more involved for anisotropic
atoms and molecules.


\section{Conclusion}

We have shown that the force between atoms with anisotropic polarizabilities
and dilute dielectrics with anisotropic permeabilities can have 
repulsive components. The 
weakly interacting configurations we considered give a
semi-quantitative understanding of the analogous
strongly interacting case. That the two-body Casimir energy
is not always monotonic in the separation is
associated to the tensorial structure of the polarizabilities and
the electromagnetic Green's dyadic.
The analysis of an anisotropic atom above a dilute dielectric plate
with a hole provides considerable insight into the 
analogous configuration involving perfect metals considered in 
\cite{Levin:2010vo}. Closed analytic solution for this configuration reveals 
torque-free points at which the interaction energy is independent of 
the orientation of the atomic polarizability. Although a single hole
in a dielectric plate does not lead to stable points,
multiple holes are expected to introduce stable points in
the potential energy surface of an oriented atom.

Unlike the scalar case, for which the irreducible three-body contribution
to Casimir energy is always positive 
\cite{Schaden:2011in,Shajesh:2011ec,Shajesh:2011se,Schaden:2011sn},
we find the electromagnetic three-body Casimir energy
can change sign. The three-body 
contributions to the energy are found to never dominate the two-body
contributions for isotropic atoms.


\section{acknowledgments}
We would like to thank the organizers of QFEXT11 for a very 
productive workshop. KVS would like to thank Iver Brevik
and Simen Ellingsen for their kind support and hospitality,
and Anand Rai for helpful discussions.
This work was supported by the National Science Foundation with 
Grant no.~PHY0902054.

\bibliography{biblio/b20110918-atoms}

\end{document}